\documentclass[10pt,number,final]{elsarticle}
\usepackage{longtable}
\usepackage{booktabs}
\usepackage[latin2]{inputenc}
\usepackage[T1]{fontenc}
\usepackage{graphicx}
\usepackage{ae}
\usepackage{aecompl}
\usepackage{setspace}
\usepackage{amsmath}
\usepackage{amssymb}
\usepackage{amsthm}
\usepackage{natbib}
\usepackage{todonotes}
\usepackage{parskip}
\usepackage{graphicx}
\usepackage{epstopdf}
\usepackage{float}
\usepackage{caption}

    \makeatletter
    \def\ps@pprintTitle{%
      \let\@oddhead\@empty
      \let\@evenhead\@empty
      \def\@oddfoot{\reset@font\hfil\thepage\hfil}
      \let\@evenfoot\@oddfoot
    }
    \makeatother

\title{Nucleotide 9-mers Characterize the Type II Diabetic Gut Metagenome}
\author[p]{Balázs Szalkai}
\ead{szalkai@pitgroup.org}
\author[p,u]{Vince Grolmusz\corref{cor1}}
\ead{grolmusz@pitgroup.org}
\cortext[cor1]{Corresponding author}
 
\address[p]{PIT Bioinformatics Group, Eötvös University, H-1117 Budapest, Hungary}
\address[u]{Uratim Ltd., H-1118 Budapest, Hungary}

\begin{document}







\vfil
\eject

\begin{abstract}

Discoveries of new biomarkers for frequently occurring diseases are of special importance in today's medicine. While fully developed type II diabetes (T2D) can be detected easily, the early identification of high risk individuals is an area of interest in T2D, too. Metagenomic analysis of the human bacterial flora has shown subtle changes in diabetic patients, but no specific microbes are known to cause or promote the disease. Moderate changes were also detected in the microbial gene composition of the metagenomes of diabetic patients, but again, no specific gene was found that is present in disease-related and missing in healthy metagenome. 
However, these fine differences in microbial taxon- and gene composition are difficult to apply as quantitative biomarkers for diagnosing or predicting type II diabetes. 
In the present work we report some nucleotide 9-mers with significantly differing frequencies in diabetic and healthy intestinal flora. To our knowledge, it is the first time such short DNA fragments have been associated with T2D. The automated, quantitative analysis of the frequencies of short nucleotide sequences seems to be more feasible than accurate phylogenetic and functional analysis, and thus it might be a promising direction of diagnostic research.
\bigskip

\end{abstract}

\maketitle

\section{Introduction}

Metagenomics \cite{national2007The} is rapidly gaining importance in clinical research \cite{Wu2010a,Qin2012a,Neu2010,Lyra2010,Cani2011,Bradley2011,Boerner2011,Amar2011}, environmental studies \cite{Fierer2012,Pandit2014,Xie2011} and biotechnology \cite{Schmeisser2007,Ferrer2007,Steele2005}. Numerous complex and reliable methods have been published for the phylogenetic identification of non-cloned short DNA reads from environmental or clinical samples, for example, the similarity-based methods MEGAN \cite{Huson2007,Huson2011,Huson2012} and MG-RAST \cite{Wilke2013,Glass2010}, the marker-gene identifying phylogenetic analyzer AMPHORA \cite{Wu2008} and its more user-friendly versions,  AMPHORA2 \cite{Wu2012b} and AmphoraNet \cite{Kerepesi2014,Kerepesi2014a}. 

These methods use multi-phase, complex approaches to retrieve phylogenetic information from the short read datasets, applying reference database operations in the process. 

Surprisingly, it was shown that simple frequency counting of nucleotides or short nucleotide sequences in the metagenomic samples may also imply phylogenetic information. 

It has been widely known for a long time that genomic AT/GC ratio is distributed in a wide range in bacterial species, and can be characteristic to some of them \cite{SUEOKA1962,Bernardi1986,Hurst2001}. The ratio is shown to be influenced by numerous environmental and metabolic factors \cite{Wu2012c} and also carries phylogenetic information.

The article \cite{Karlin1997} reports differences in di- and tetranucleotide frequencies among numerous bacterial species, and examines the possible application of these signatures in molecular phylogeny. 

Tetranucleotide sequence frequencies were applied in supervised and unsupervised phylogenetic classification, or ``binning'' in \cite{Saeed2009}.

The work \cite{Krause2008} applies conserved gene fragments, each encoding several dozens of amino acids, identified from the Pfam database \cite{Sonnhammer1997a}. The fragments are called ``environmental gene tags'', and are used successfully for phylogenetic binning in \cite{Krause2008}.

The study of \cite{Qin2012a} investigated the differences in gut metagenomes of diabetic and healthy subjects. The metagenomes were {\em de novo} assembled,  and the bacterial genes were mapped to a metagenomic gene catalog. Genes related to oxidative stress response were found more abundant in the samples originating from diabetic subjects. Additionally, moderate changes in intestinal bacterial composition were detected, but no specific microbes were associated with the metagenomes of the type II diabetes (T2D) patients. 

After a very complex selection and filtering process, genome-specific nucleotide markers of length 50 were identified in \cite{Tu2014}. The markers were applied for strain/species identification, and also as markers for microbial species that might play a role in T2D and obesity in the data set of \cite{Qin2012a}.

Here we describe a very simple and straightforward approach for finding short nucleotide sequences whose frequencies significantly differ in T2D and healthy metagenomes of the dataset of \cite{Qin2012a}. We identify several nucleotide 9-mers that may serve as quantitative biomarkers of the pre-diabetic state in the future. To our knowledge, such short sequences have never been found to characterize T2D or any other disease.

We need to clarify that we do not state that the identified 9-mers will generally be applicable as biomarkers for diabetes for all human populations. We believe that ``enterotype-specific'' \cite{Arumugam2011} quantitative biomarkers could be found for each enterotypes by exhaustive searches described in the Methods section, and those enterotype-specific biomarkers could serve as predictors of type 2 diabetes mellitus.

\section{Discussion and Results}

Our results are summarized on Table 1 and on Figure 1. Table 1 contains 20 7-, 8- and 9-mers of the highest statistical significance, distinguishing between the diabetic and non-diabetic metagenomes of the study \cite{Qin2012a}. 

Table 1 was prepared without considering complementarities between the short nucleotide sequences. Therefore, the complements found with very close frequencies and statistical parameters independently verify our results. It is easy to recognize in Table 1 that TGTGGTA and TACCACA are exact complements. The complement of TCCACAT, ATGTGGA, is almost the prefix of ATGTGGTAC. The complement of TGTGGTACT (line 3) is again the exact complement of AGTACCACA (line 6), just to mention some of the complementarities in the table.

Figure 1 gives the empirical cumulative distribution functions of the frequency of 9-mer TGTGGTGTA in the diabetic and in the non-diabetic samples. The difference between the expected values (means) of the two distribution is obvious on the figure and is quantified statistically in Table 1. 

We also searched for short nucleotide sequences characterizing lean/obese and male/female individuals in the dataset of \cite{Qin2012a}.
Only one short sequence passed the statistical significance bound in the lean/obese search, and none in the male/female search (c.f. Table S1 and S2 and Figure S1 in the Appendix).

The source of the bias in short nucleotide sequence frequencies is most probably due to the difference in the gene- and species composition of diabetic and healthy metagenomes, found in \cite{Qin2012a,Tu2014}. These frequencies could be measured and evaluated more easily than the much more involved characteristics found in \cite{Qin2012a,Tu2014}.

	\section{Materials and Methods}
	
	Our data source was the set of metagenomes of 345 Chinese subjects, collected by Qin et al. \cite{Qin2012a} and deposited in the Sequence Read Archive (http://www.ncbi.nlm.nih.gov/sra) under accession numbers SRA045646 (145 subjects) and SRA050230 (225 subjects). The assembled data was downloaded from the GigaScience database, GigaDB at http://dx.doi.org/10.5524/100036.    
	
	We considered all the possible DNA sequences of length at most 9 (this means over 300,000 possible sequences). For each sequence, we counted the number of exact matches in each raw metagenome. Our aim was to determine whether there are any short DNA fragments whose frequencies differ for diabetic/non-diabetic, lean/obese or female/male individuals.
	
	We first defined the frequency of a short DNA fragment for a given metagenome as the number of occurrences (exact matches), divided by the total size, measured in base-pairs (bp), of the metagenome. Additionally -- to account for minor mutations --  we also included those sequences in the counting process that differed by only one nucleotide, but these were considered with half a weight. So, for example, the final {\em frequency} of the sequence AAA included not only how many times the sequence AAA occurs in a specific metagenome, but also how many times AAG, CAA, ATA, ... occur in that metagenome, except that the number of occurrences for these related DNA fragments was divided by two.
	
	Let $\ell_M$ denote the length in base-pairs (bp) of a metagenome $M$. Let $d(s, t)$ be the number of mismatches between the two sequences of same length, $s$ and $t$ (also called the Hamming distance). Let $k_M(s)$ denote the number of exact matches of sequence $s$ in metagenome $M$. Then $f_M(s)$ (the frequency of sequence $s$ with respect to metagenome $M$) is defined by the formula
	
	\[
	f_M(s) = \frac{1}{\ell_M}\left( k_M(s) + \frac{1}{2} \sum_{d(s, t) = 1} k_M(t)\right).
	\]
	
	This approach (counting some non-exact matches as well, but with the half the weight) yielded statistically better results when compared to the original, stricter counting process, which only allowed exact matches.
	
	We developed C++ programs for counting the fragments and analyzing the results. Several partitions on the set of subjects were analyzed, by dividing them into two groups by different attributes: diabetic/non-diabetic, lean/obese and female/male. Our aim was to look for short DNA sequences whose mean frequency differs for the two groups. 
	
	To achieve this, first we calculated $f_M(s)$ for each raw/assembled metagenome $M$ and each short DNA sequence $s$ of length $\ell_s \leq 9$. Then, for each $s$ we calculated a $p$-value using Welch's t-test, which showed whether the frequency $s$ is the same in the two groups (i.e., $p$ is large) or differs significantly (i.e., $p \leq 0.05$).

	\begin{figure}[H]
		\centering
		\includegraphics[width=0.75\textwidth]{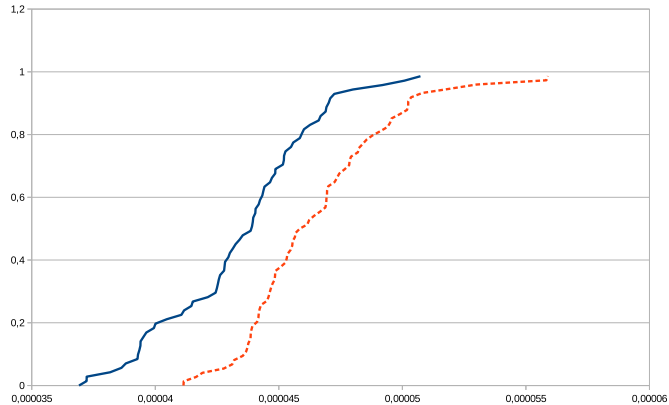}
		\caption{Empirical cumulative distribution function of the frequency of 9-mer TGTGGTGTA (solid: diabetic, dashed: non-diabetic). For every value $x$ the  curves demonstrate the diabetic (solid line) and non-diabetic (dashed line) fraction of metagenomes with TGTGGTGTA frequency of at most $x$. For example, for $x=0.000045$, 70\% of the diabetic samples have the TGTGGTGTA frequency less than $x$, while only 38\% of the non-diabetic samples have that frequency less than $x$. Further empirical cumulative distribution functions are given in the Appendix. }
	\end{figure}
	
	Since this was done for each short DNA fragment, the number of total statistical tests done for a given division of subjects was equal to the number of possible $s$ DNA sequences of length at most 9. As this is more than 300,000, there was a high probability that one of the tests would yield a very low $p$-value but the large measured difference of means would be in fact due to mere chance.
	
	Therefore we utilized a two-step hypothesis testing procedure. First we computed the $p$-values for Study 1 (with 145 subjects, SRA accession number SRA045646) only, which now became our {\em training set}. Then we sorted the possible $s$ sequence candidates by $p$-value ascending, and chose those 20 sequences which had the lowest $p$-value. These were those sequences which showed promise that their frequency might differ significantly between diabetic/non-diabetic, lean/obese and female/male individuals, depending our current partitioning of the subjects. Then we tested these selected sequences (and corresponding statistical hypotheses) on the {\em holdout set}, which was the collection of metagenomes from Study 2 (SRA accession number SRA050230, 225 subjects). On this set we performed only those 20 tests which qualified in the first round, which again yielded a second $p$-value for each of the 20 DNA sequences.
	
		{\small

			\begin{longtable}{l | ccccc}\label{1}
				Fragment & Diabetic & Non-diabetic & p (training set) & p (holdout set) & p (corrected) \\ 
				TGTGGTGTA & 4.475e-05 & 4.713e-05 & \textbf{7.8e-09} & 0.000296 & 0.005928 \\ 
				TGTGCTATC & 4.346e-05 & 4.549e-05 & \textbf{1.871e-08} & 0.001764 & 0.033518 \\ 
				TGTGGTACT & 4.009e-05 & 4.164e-05 & \textbf{9.508e-10} & 0.001929 & 0.034726 \\ 
				TGTGGTA & 0.0006214 & 0.0006428 & \textbf{1.397e-08} & 0.001937 & 0.032922 \\ 
				TGTGGTACA & 4.672e-05 & 4.875e-05 & \textbf{2.973e-08} & 0.002098 & 0.033564 \\ 
				AGTACCACA & 4.096e-05 & 4.242e-05 & \textbf{2.152e-08} & 0.002246 & 0.033686 \\ 
				CCATCTGT & 0.0002318 & 0.0002424 & \textbf{2.138e-08} & 0.003092 & 0.043284 \\ 
				TGCCACATA & 5.811e-05 & 6.126e-05 & 6.417e-09 & 0.004678 & 0.060818 \\ 
				TGTGGTATG & 4.813e-05 & 5.04e-05 & 9.19e-09 & 0.004925 & 0.059099 \\ 
				TACCACA & 0.0006332 & 0.0006531 & 3.377e-08 & 0.004999 & 0.054987 \\ 
				TGTGGAGAT & 6.544e-05 & 6.788e-05 & 1.523e-08 & 0.008901 & 0.089010 \\ 
				TGTGGTATC & 5.035e-05 & 5.248e-05 & 1.492e-08 & 0.011902 & 0.107118 \\ 
				ATGGTCTGT & 5.845e-05 & 6.071e-05 & 1.291e-08 & 0.012383 & 0.099068 \\ 
				GTACCACAT & 4.179e-05 & 4.311e-05 & 1.055e-08 & 0.012814 & 0.089698 \\ 
				CCACATACT & 5.127e-05 & 5.348e-05 & 2.436e-08 & 0.014294 & 0.085764 \\ 
				ATGTGGTAC & 4.135e-05 & 4.266e-05 & 9.495e-09 & 0.024340 & 0.121702 \\ 
				TCTCCACAT & 6.968e-05 & 7.255e-05 & 1.582e-08 & 0.074780 & 0.299121 \\ 
				ATCTCCACA & 6.615e-05 & 6.84e-05 & 5.427e-09 & 0.078516 & 0.235547 \\ 
				CTCCACATA & 5.578e-05 & 5.753e-05 & 2.018e-08 & 0.257111 & 0.514221 \\ 
				TCCACAT & 0.0008132 & 0.0008294 & 1.919e-08 & 0.266428 & 0.266428 \\ 
				\caption{Frequencies of 7-, 8- and 9-mers in diabetic vs. non-diabetic samples with the highest significance (training set: Study 1, holdout set: Study 2). The training set p-values are highlighted for the statistically significant multimers ($p < 0.05$ with Holm-Bonferroni correction). It is easy to recognize that TGTGGTA and TACCACA are exact complements. The complement of TCCACAT, ATGTGGA, is almost the prefix of ATGTGGTAC. 9-mer TGTGGTACT (line 3) is the exact complement of AGTACCACA (line 6). One can find further complementarities in the table. These independently found complements with very close frequencies and p-values strengthen our findings. More tables (for lean-obese and female-male distributions) are given in the Appendix.}
			\end{longtable}}

	Then the Holm-Bonferroni correction was used to determine which of the sequences had a significantly different frequency among the two groups. This correction algorithm effectively calculates an upper bound for a $p$-value which takes the fact that we performed multiple (i.e., 20) statistical tests into account. Since the frequencies in the second study are independent from those in the first study, the first one is indeed a suitable training set for the model, and we can safely ignore that we performed over 300,000 statistical tests on the first study, since we use only the tests on the holdout set to make predictions.
	
	We have applied the raw, unassembled metagenomes from Study 1 and Study 2 to look for short marker sequences of diabetes. 
	
	Unfortunately, there was not enough information available to us to determine which subjects of Study 2 are lean/obese or female/male. Thus we had to use the available assembled metagenomes in Study 1 to look for marker fragments for sex and obesity. We partitioned the assembled metagenomes of the first study into two ``random'' groups: one of the groups consisted of those individuals with an odd subject ID, and the other group contained those with an even ID. One of these was the training set and the other became the holdout set, i.e. they took the role of Study 1 and Study 2 for the lean/obese and female/male classifications (Tables S1 and S2 in the Appendix).
	
	 One sequence passed the significance threshold for the lean/obese division, and none of the short sequences had a significant difference of frequency between the two sexes.

	\section{References}
	

\newpage

\section{Appendix}

{\small

	\begin{longtable}{l | ccccc}\label{S1}
		Fragment & Lean & Obese & p (training set) & p (holdout set) & p (corrected) \\ 
		CTCGTGACA & 2.002e-05 & 1.901e-05 & \textbf{0.002091} & 0.001443 & 0.028859 \\ 
		CTCGATTGT & 2.848e-05 & 2.727e-05 & 0.002945 & 0.004539 & 0.086232 \\ 
		TGTCGACTG & 2.459e-05 & 2.3e-05 & 0.0009184 & 0.005781 & 0.104062 \\ 
		ACACTCGAG & 1.126e-05 & 1.025e-05 & 0.001831 & 0.006911 & 0.117490 \\ 
		CTCGAGTGT & 1.127e-05 & 1.025e-05 & 0.002036 & 0.012364 & 0.197821 \\ 
		TGTGACTCG & 1.354e-05 & 1.291e-05 & 0.002158 & 0.014499 & 0.217484 \\ 
		ATGTGAGGC & 2.35e-05 & 2.255e-05 & 0.001805 & 0.016175 & 0.226446 \\ 
		GTGCCTCTC & 2.382e-05 & 2.26e-05 & 0.002931 & 0.019559 & 0.254270 \\ 
		GGCTCACTC & 1.817e-05 & 1.722e-05 & 0.003306 & 0.031810 & 0.381714 \\ 
		CGAGTGAGA & 1.858e-05 & 1.786e-05 & 0.003293 & 0.036067 & 0.396741 \\ 
		CACTCGAGG & 1.205e-05 & 1.087e-05 & 0.003403 & 0.061201 & 0.612006 \\ 
		GAGTGAGCT & 2.149e-05 & 2.059e-05 & 0.003223 & 0.062982 & 0.566836 \\ 
		CTCGACTGT & 2.062e-05 & 1.954e-05 & 0.003178 & 0.071181 & 0.569449 \\ 
		CTGTCGTGT & 2.723e-05 & 2.629e-05 & 0.00301 & 0.077670 & 0.543687 \\ 
		TGTGGTTGA & 5.722e-05 & 5.518e-05 & 0.002553 & 0.121549 & 0.729294 \\ 
		CACTCGTGG & 1.633e-05 & 1.524e-05 & 0.002677 & 0.130222 & 0.651109 \\ 
		TCACCATGT & 4.975e-05 & 4.83e-05 & 0.003499 & 0.283407 & 1.133628 \\ 
		TCTAGCCTG & 1.786e-05 & 1.729e-05 & 0.003271 & 0.561284 & 1.683851 \\ 
		AACAGCCAC & 5.328e-05 & 5.22e-05 & 0.002606 & 0.697098 & 1.394196 \\ 
		CTAGCTGTC & 2.083e-05 & 2.036e-05 & 0.001805 & 0.882905 & 0.882905 \\ 
		\caption*{Table S1: Frequencies of ninemers of in lean vs. obese samples with the highest significance (training and holdout sets: two halves of Study 1). The boldface number in column 4 denotes the significant difference by Holm-Bonferroni corrections, shown in the last column.}
	\end{longtable}
	
	\newpage
	
	\begin{longtable}{l | ccccc}\label{4}
		Fragment & Male & Female & p (training set) & p (holdout set) & p (corrected) \\ 
		TAGTACTGG & 2.748e-05 & 2.854e-05 & 0.006019 & 0.174548 & 3.490951 \\ 
		TTCATAGGG & 3.385e-05 & 3.479e-05 & 0.0005157 & 0.305204 & 5.798868 \\ 
		AGTCTCAGG & 2.314e-05 & 2.229e-05 & 0.007333 & 0.353644 & 6.365594 \\ 
		GATGTGTCT & 3.878e-05 & 3.841e-05 & 0.006985 & 0.452399 & 7.690776 \\ 
		GTCTCACAC & 1.635e-05 & 1.594e-05 & 0.00236 & 0.495140 & 7.922244 \\ 
		CTCAGTCT & 0.0001047 & 0.0001014 & 0.006424 & 0.512597 & 7.688961 \\ 
		CATGTAACC & 2.969e-05 & 2.932e-05 & 0.001608 & 0.515833 & 7.221663 \\ 
		GCTTCAGAC & 4.097e-05 & 3.98e-05 & 0.006813 & 0.546829 & 7.108781 \\ 
		CTCTAACAC & 2.147e-05 & 2.098e-05 & 0.006313 & 0.578498 & 6.941978 \\ 
		ACAGACTCA & 3.893e-05 & 3.82e-05 & 0.007392 & 0.582096 & 6.403058 \\ 
		GGTCAATTC & 4.215e-05 & 4.266e-05 & 0.006413 & 0.595760 & 5.957600 \\ 
		TGTGAGTCT & 2.247e-05 & 2.204e-05 & 0.007573 & 0.618236 & 5.564123 \\ 
		CAGACTCAT & 4.513e-05 & 4.426e-05 & 0.007669 & 0.619291 & 4.954327 \\ 
		GTGTTAGAC & 1.626e-05 & 1.596e-05 & 0.004958 & 0.625175 & 4.376228 \\ 
		ACCTCTGTC & 4.032e-05 & 3.957e-05 & 0.005543 & 0.729250 & 4.375499 \\ 
		GTCTAACAC & 1.634e-05 & 1.596e-05 & 0.002582 & 0.752233 & 3.761164 \\ 
		AGGATGTGT & 4.805e-05 & 4.725e-05 & 0.001627 & 0.795980 & 3.183920 \\ 
		TCTCCTCAA & 5.775e-05 & 5.652e-05 & 0.006681 & 0.909561 & 2.728684 \\ 
		TCTCAGTCT & 3.361e-05 & 3.263e-05 & 0.004097 & 0.945748 & 1.891496 \\ 
		GGTGTGTCT & 2.855e-05 & 2.794e-05 & 0.005231 & 0.949554 & 0.949554 \\ 
		\caption*{Table S2: Frequencies of ninemers and an eightmer in female vs. male samples with the highest significance (training and holdout sets: two halves of Study 1). After the very strict Holm-Bonferroni corrections, no significant differences were found. }
	\end{longtable}
	
}

\newpage

\bigskip

\begin{figure}[H]
	\centering
	\includegraphics[width=0.75\textwidth]{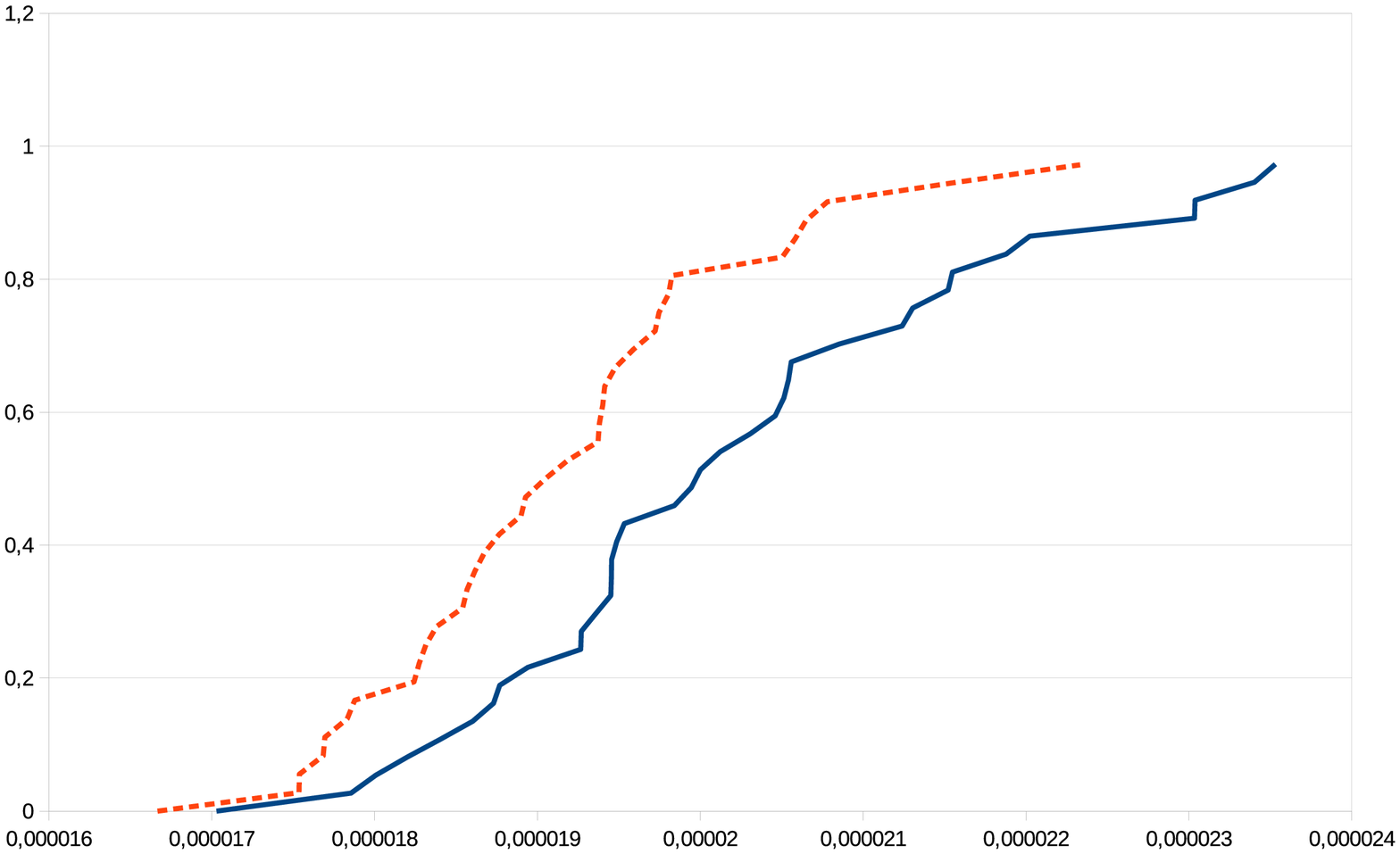}
	\captionsetup{labelformat=empty}
	\caption*{Figure S1: Empirical cumulative distribution function of the frequency of fragment CTCGTGACA (solid: lean, dashed: obese)}
\end{figure}

\end{document}